\documentclass[12pt]{iopart}

\usepackage{iopams}

\usepackage{graphicx}
\begin{document}

\title[An Infinite-Period Phase Transition versus Nucleation in a
  Stochastic Model]{An infinite-period phase transition versus
  nucleation in a stochastic model of collective oscillations}

\author{Vladimir R. V. Assis$^{1,2,3}$, Mauro Copelli$^3$ and Ronald Dickman$^4$}
\address{$^1$Departamento de F{\'\i}sica, Universidade Estadual de Feira de Santana,
  44031-460, Feira de Santana, Bahia, Brazil}
\address{$^2$Centro Brasileiro de Pesquisas F{\'\i}sicas, R. Dr. Xavier Sigaud,
  150,  22290-180 Rio de Janeiro-RJ, Brazil}
\address{$^3$Departamento de F{\'\i}sica, Universidade Federal de Pernambuco,
  50670-901 Recife, PE, Brazil}
\address{$^4$Departamento de F\'{\i}sica, Instituto de
Ci\^encias Exatas, and National Institute of Science and Technology for Complex Systems,
Universidade Federal de Minas Gerais
C.P. 702, 30123-970, Belo Horizonte, MG, Brazil}
\ead{vladimir@cbpf.br and mcopelli@df.ufpe.br and dickman@fisica.ufmg.br}

\begin{abstract}
  
  A lattice model of three-state stochastic phase-coupled oscillators
  has been shown by Wood {\it et al.\/} (2006 {\it Phys. Rev. Lett.\/}
  {\bf 96} 145701) to exhibit a phase transition at a critical value
  of the coupling parameter, leading to stable global oscillations. We
  show that, in the complete graph version of the model, upon further
  increase in the coupling, the average frequency of collective
  oscillations decreases until an infinite-period (IP) phase
  transition occurs, at which point collective oscillations
  cease. Above this second critical point, a macroscopic fraction of
  the oscillators spend most of the time in one of the three states,
  yielding a prototypical nonequilibrium example (without an
  equilibrium counterpart) in which discrete rotational ($C_3$)
  symmetry is spontaneously broken, in the absence of any absorbing
  state. Simulation results and nucleation arguments strongly suggest
  that the IP phase transition does not occur on finite-dimensional
  lattices with short-range interactions.
\end{abstract}

\maketitle

\tableofcontents

\section{Introduction}

Models of coupled oscillators exhibit a surprising range of
symmetry-breaking transitions to a globally synchronized state.  In
the paradigmatic Kuramoto model, for instance, oscillators with
intrinsic frequencies $\omega_i$ coupled via their continuous phases
$\theta_i$ can exhibit stable collective oscillations, with breaking
of time translation
symmetry~\cite{Kuramoto84,Strogatz93,Strogatz00,StrogatzSync,Pikovsky01}. When
$\omega_i=0$, the Kuramoto model acquires an additional $\theta_i \to
-\theta_i$ reflection symmetry, which can also be broken at the onset
of synchronization, yielding surprising response
properties~\cite{Reimann99b}.

Within the class of discrete-phase models, the paper-scissors-stone
game is an example of a system with three absorbing states which can
present either global oscillations or spontaneous breaking of discrete
rotational ($C_3$)
symmetry~\cite{Tainaka88,Tainaka89,Tainaka91,Itoh94,Tainaka94}.  More
recently, Wood, Van den Broeck, Kawai and Lindenberg proposed a family
of models of phase-coupled three-state stochastic oscillators which
can undergo a phase transition to a synchronized
state~\cite{Wood06a,Wood06b,Wood07a,Wood07b} for sufficiently strong
coupling. From here on each of them is called Wood {\it et al.}'s
cyclic model~(WCM). Although the WCM also has $C_3$ symmetry, it has
no absorbing state.

As noted previously~\cite{Wood07a}, the first WCM~\cite{Wood06a} has
an unusual behavior after the phase transition to a synchronized
state. As the phase coupling is further increased, collective
oscillations slow down while oscillators remain partially
synchronized. Here we have a closer look at this phenomenon, in
particular addressing whether or not there is a second phase
transition with breaking of $C_3$ symmetry.

The paper is organized as follows. In section~\ref{model} we review
the WCM and the essentials of the known transition to a synchronized
phase. Our results regarding a putative second phase transition are
described in section~\ref{results}. We conclude in
section~\ref{conclusions}.
\begin{figure}[!b]%
  \begin{center}%
    \includegraphics[width=0.9\columnwidth,angle=0]{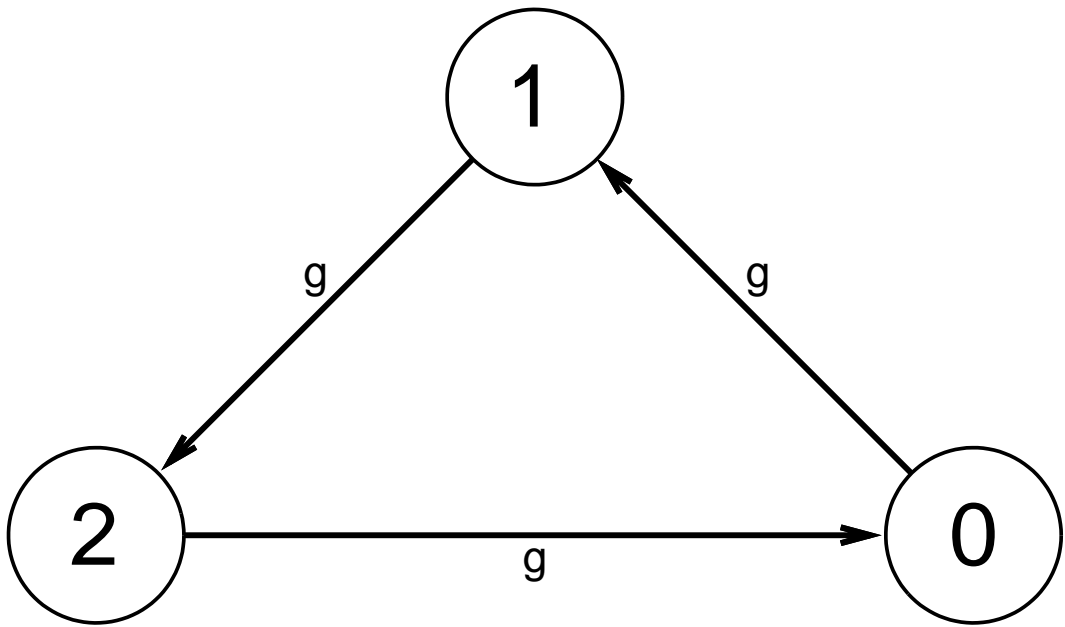}%
    \caption{\label{fig:taxas} Transition rates for an isolated unit.}%
  \end{center}%
\end{figure}%

\begin{figure}
\includegraphics[width=0.9\textwidth,angle=0]{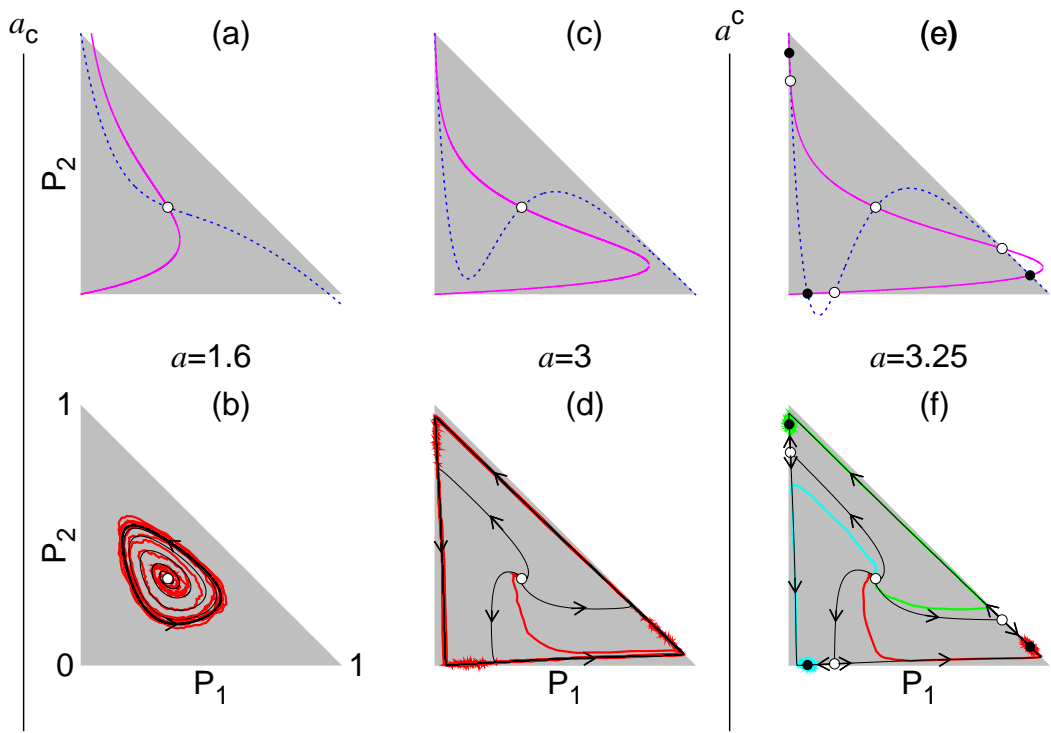}
\begin{center}
\caption{\label{fig:phasediagram}(Color online)~Phase space~$(P_1,
  P_2)$. Grey triangles correspond to the allowed region~$0 \leq P_2
  \leq 1-P_1$. $a$ increases from left to right. In the upper panels,
  dashed~(solid) lines denote the
  nullclines~$\dot{P}_1=0$~($\dot{P}_2=0$). Lower panels show
  trajectories from complete graph simulations with~$N=5000$~(thick
  lines) and solutions of the mean-field equations~(thin lines with
  arrows). Panel~(a) shows that~$\vec{P}_{1/3}^*$ is the only fixed
  point in the system for~$a=1.6 > a_c$. (b)~$\vec{P}_{1/3}^*$ is
  unstable, with trajectories converging to a stable limit cycle,
  which corresponds to sustained global oscillations (see also
  Fig.~\ref{complete_graph}a). (c)~For~$a=3$, the nullclines approach
  one another in three additional positions, thereby creating three
  symmetric saddle-node ghosts. (d)~The limit cycle is almost as large
  as the perimeter of the allowed region. Note the overcrowding of
  lines near the ghosts, with consequent anharmonicity in the time
  series~(Fig.~\ref{complete_graph}b). (e)~For~$a=3.25$ there are six
  additional fixed points which are born simultaneously in three
  saddle-node bifurcations at~$a=a^c$. (f)~In this case, if the system
  starts from the symmetric unstable fixed point~$\vec{P}_{1/3}^*$,
  fluctuations determine to which stable fixed point it
  goes. Closed~(open) symbols denote stable~(unstable) fixed points.}
\end{center}
\end{figure}

\begin{figure}
\includegraphics[width=0.9\columnwidth,angle=0]{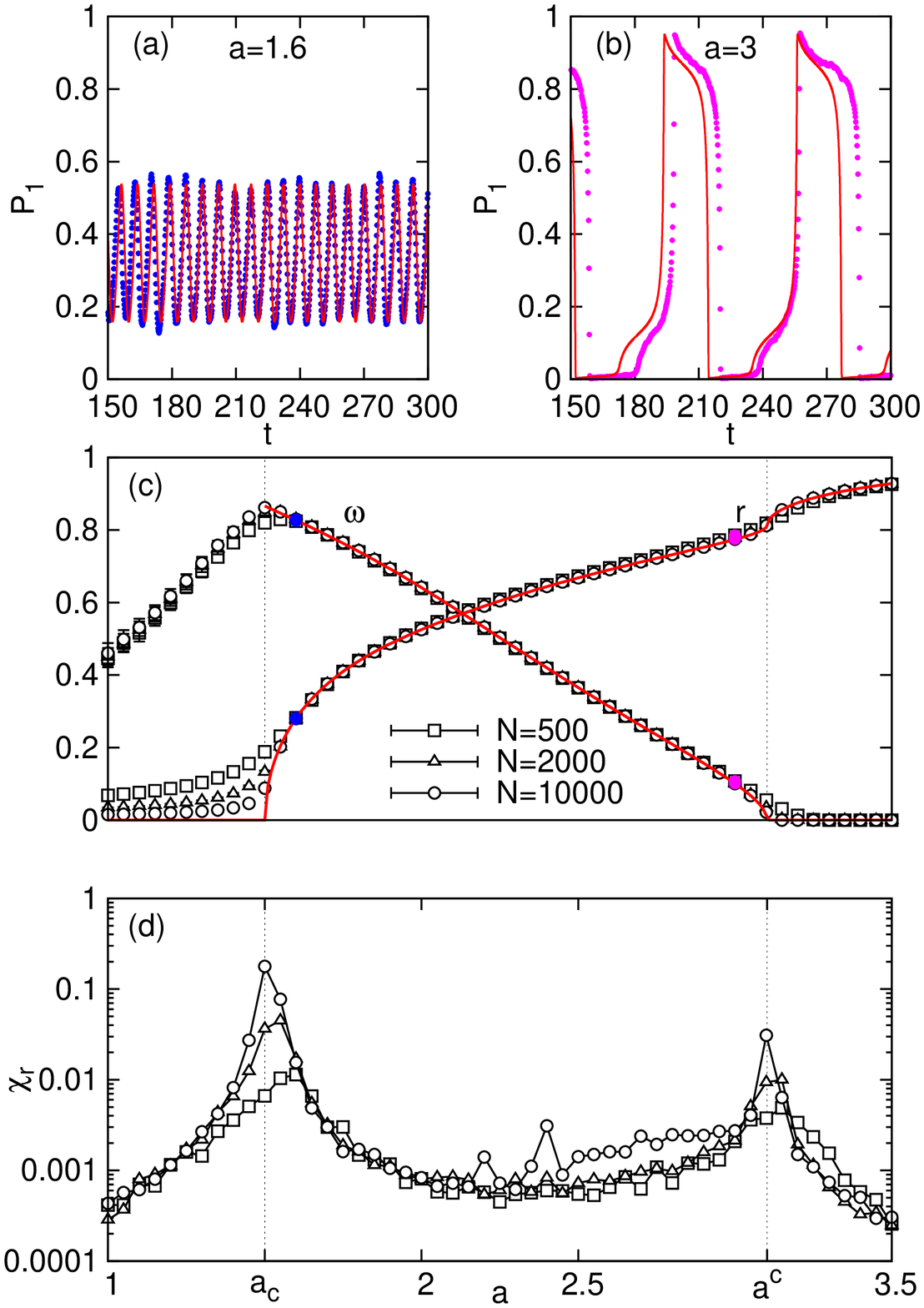}
\begin{center}
\caption{\label{complete_graph} (Color online)~Panels~(a) and~(b) show
  the evolution of~$P_1$ for~$a=1.6$ and~$a=3$, respectively. Points
  represent simulations for complete graph with~$N=10000$ whereas
  lines show the mean-field solution. (c)~Dependence of $r$ and
  $\omega$ on $a$, exhibiting the two phase transitions. (d)~$\chi_r$
  versus~$a$ showing peaks at the transitions [system sizes as
    in~(c)].}
\end{center}
\end{figure}

\section{\label{model} Model}

In the~WCM, the phase~$\phi_x$ at site~$x$~($x=1,\ldots,N$) can take
one of three values:~$\phi_x = j_x2\pi/3$, where~$j_x \in \{0,1,2\}$
(for simplicity, from here on we employ~$x$ to denote sites and $j$ to
denote states --- always modulo 3). The only allowed transitions are
those from~$j$ to~$j+1$ (see Fig.~\ref{fig:taxas}), with transition
rates
\begin{equation}
\label{eq:gj}
g_{j, j+1} = g \exp \left[ \frac {a \left( N_{j+1} - N_j \right)} {z} \right],
\end{equation}
where~$g$ is a constant (which can be set to unity without loss of
generality),~$a$ is the coupling parameter,~$N_j$ is the number of
nearest neighbors in state~$j$, and~$z$ is the number of nearest
neighbors. The rates are invariant under cyclic permutation of the
indices, i.e., belong to the symmetry group~$C_3$ of discrete
rotations, and strongly violate the detailed balance condition.

The mean-field approximation is obtained by replacing~$N_j/z$ in the
argument of the exponential of Eq.~\ref{eq:gj} with the corresponding
probability,~$P_j$, leading to the following set of equations:
\begin{equation}
\label{eq:MF}
\dot P_j = g_{j-1,j}P_{j-1}-g_{j,j+1}P_{j}\;,
\end{equation}
where now~$g_{j,j+1} = g \exp \left[ a \left( P_{j+1} - P_j \right)
  \right]$. These also represent the equations for a complete graph in
the limit~$N\to\infty$: since~$z=N-1$, we can replace~$N_j/N$
with~$P_j$ in the infinite-size limit. Normalization reduces these
equations to a two-dimensional flow in the~$(P_1,P_2)$
plane~(Fig.~\ref{fig:phasediagram}).

Starting from the complex order parameter,
\begin{equation}
v \equiv \frac{1}{N} \sum^{N}_{x=1} e^{i\phi_x}
\end{equation}
one can define quantities that characterize the collective behavior of
the system. A configuration with~$v \neq 0$ has unequal numbers of
sites in the three states, which led Kuramoto to propose
\begin{equation}
r \equiv \left<\left< |v| \right>_t\right>_s
\end{equation}
as the order parameter for
synchronization~\cite{Kuramoto84,Strogatz00,Wood06a}, where $\left<
\right>_t$ denotes a time average over a single realization (in the
stationary state), and $\left< \right>_s$ an average over independent
realizations. Note that $r > 0$ is consistent with, but does not
necessarily imply, globally synchronized oscillations.  The latter is
characterized by a periodically varying phase
of~$v$~\cite{Kuramoto92a,Ohta08,ShinomotoKuramoto86,Rozenblit11}.

In the mean-field equations~(\ref{eq:MF}), the transition to the
synchronized regime is associated with a supercritical Hopf
bifurcation at $a=a_c=1.5$.  The trivial fixed point
$\vec{P}_{1/3}^*\equiv (P_1=1/3,P_2=1/3)$ loses stability at $a=a_c$,
and a limit cycle encircling this point appears
(Fig.~\ref{fig:phasediagram}a-b)~\cite{Wood06a}. For $a \gtrsim a_c$,
sustained oscillations in $P_j$ characterize synchronization among the
oscillators (Fig.~\ref{complete_graph}a). Correspondingly, $r$ grows
continuously~\mbox{$\sim (a-a_c)^\beta$} at the transition
(Fig.~\ref{complete_graph}c), with a mean-field exponent~\mbox{$\beta
  = 1/2$}~\cite{Wood06a}.  The scaled variance $\chi_r \equiv L^d
\left[ \left< \left< |v| \right>^2_t \right>_s - r^2 \right]$ diverges
with the system size at criticality, as shown for simulations of
complete graphs using~$L^d\equiv N$ in Fig.~\ref{complete_graph}d.

Note that the phase transition to stable collective oscillations is
associated with the breaking of the continuous time-translation
symmetry: the time series~$P_0(t)$,~$P_1(t)$ and~$P_2(t)$ become
periodic for~$a\gtrsim a_c$. But since they are statistically
identical, except for a phase,~$C_3$ symmetry still
holds. Increasing~$a$ above~$a_c$ enhances synchronization among the
oscillators, leading to increasing oscillation amplitudes, as shown in
Figs.~\ref{fig:phasediagram}d and~\ref{complete_graph}b.

\section{\label{results} The Second Phase Transition}

\subsection{Mean-field theory and the complete graph}

Wood~{\it et al.} found that the increasing amplitude of oscillation
is accompanied by a decreasing angular frequency~$\omega =
2\pi/\langle \tau \rangle$, where~$\langle \tau \rangle$ is the mean
time between peaks in~$P_k$~(Figs.~\ref{complete_graph}a-c). This can
be understood qualitatively from the exponential dependence of the
rates of Eq.~\ref{eq:gj} on the neighbor fractions: when a state is
highly populated, the rate at which oscillators leave this state
becomes very small.  In addition, the rate at which a highly populated
state attracts oscillators from its predecessor is also very high.

We study the effect of a further increase in the coupling~$a$, beyond
the range of values investigated in Ref.~\cite{Wood06a}, using
mean-field theory and analysis on a complete graph of~$N$ sites. Note
that the state of the latter is completely specified by two integer
random variables,~$N_1$ and~$N_2$.  We study the complete graph
(for~$N$ up to 1000) via exact (numerical) stationary solution of the
master equation~\cite{Dickman02} and (for~$N$ ranging from 10 to
10$^4$) via event-driven Monte Carlo simulations.  The results of the
two approaches are fully consistent.

In mean-field theory, when~$a$ reaches an upper critical
value~$a^c\simeq 3.102\, 439\, 915\, 64$, three symmetric saddle-node
bifurcations occur simultaneously, and the period of the collective
oscillations diverges. Above~$a^c$, there are three symmetric
attractors in the system, and 3-fold rotational~($C_3$) symmetry is
spontaneously broken: Figure~\ref{fig:phasediagram}f illustrates how
the final fate of three different trajectories starting from the same
initial conditions is determined by fluctuations. Moreover, the
transition can be considered reentrant, in the sense that
time-translation symmetry, which had been broken at~$a_c$, is restored
at~$a^c$~(although, unlike other cases seen in the
literature~\cite{VandenBroeck97b,Kawai04}, the phase above $a^c$ is
not exactly the same as that below $a_c$, because of $C_3$ symmetry
breaking).

Analogously to what occurs in a condensation or ferromagnetic phase
transition~\cite{Huang}, the freezing of the relative occupancy of
each state does not imply that individual oscillators freeze as well.
The frequencies of individual oscillators do decrease with
increasing~$a$, but only vanish in the limit~$a\to\infty$, when one of
the states is fully occupied.

It is convenient to define an order parameter that is identically
zero~(in the infinite-size limit) for~$a>a^c$. Since the angular
frequency~$\omega$ does not fulfill the usual requirements for an
order parameter~\cite{Plischke&Bergersen1994}, we propose
\begin{equation}
  \label{eq:psi}
  |\psi| \equiv \frac{1}{N}\left|
\sum_{x=1}^N \left(
\delta_{0,j_x} + e^{2\pi i/3}\delta_{1,j_x} + e^{-2\pi i/3}\delta_{2,j_x}
\right) \gamma_x
\right|\; ,
\end{equation}
where~$\delta$ is the Kronecker delta and~$\gamma_x \equiv g_{j_x,j_x
  + 1}$ is the transition rate at site~$x$ (see
Eq.~\ref{eq:gj}). Thus~$|\psi|$ reflects not only the configuration,
but the rate at which the latter evolves.

On the complete graph,~$\gamma_x$ is the same for all sites~$x$ in the
same state~$j$. Denoting this rate by~$\gamma_j$, the order parameter
can be rewritten for mean-field~(MF) analysis as
\begin{equation}
  \label{eq:psiMF}
  |\psi|^2 \stackrel{MF}{=} \left[\sum_{j=0}^2 P_j^2\gamma_j^2 \right] -
  P_0\gamma_0P_1\gamma_1 - P_1\gamma_1P_2\gamma_2 -
  P_2\gamma_2P_0\gamma_0 \; .
\end{equation}
Note that both the disordered phase~($a< a_c$) as well as the frozen
phase~($a>a^c$) have stable fixed points, so~$P^*_j\gamma_j =
P^*_{j+1}\gamma_{j+1}$ from Eq.~\ref{eq:MF}. This in turn renders
$|\psi|=0$ in Eq.~\ref{eq:psiMF} for both cases [a similar line of
  reasoning can be applied directly to Eq.~(\ref{eq:psi})].

\begin{figure}%
  \begin{center}%
    \includegraphics[width=\columnwidth,angle=0]{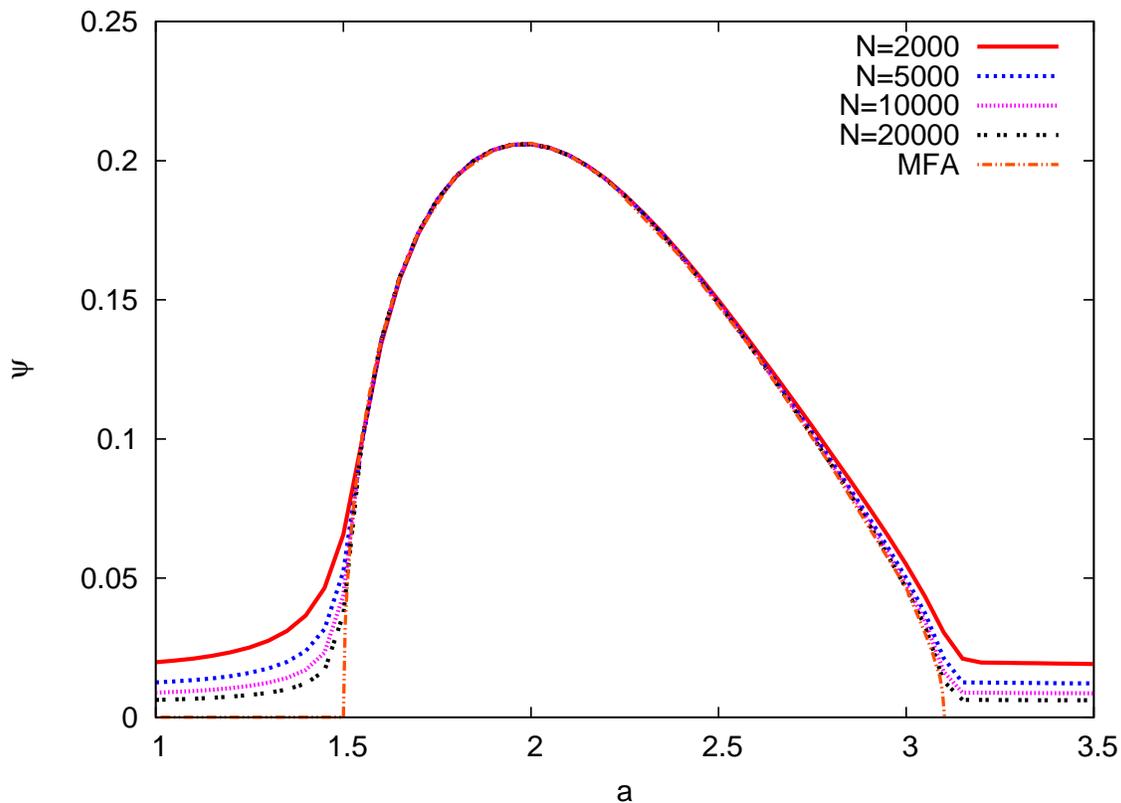}
    \caption{\label{fig:psi} Order parameter $\langle |\psi| \rangle$ as a
      function of coupling $a$ in mean-field theory and on the complete graph, for
      sizes as indicated.
       }%
  \end{center}%
\end{figure}%

\begin{figure}%
  \begin{center}%
    \includegraphics[width=1.2\columnwidth,angle=0]{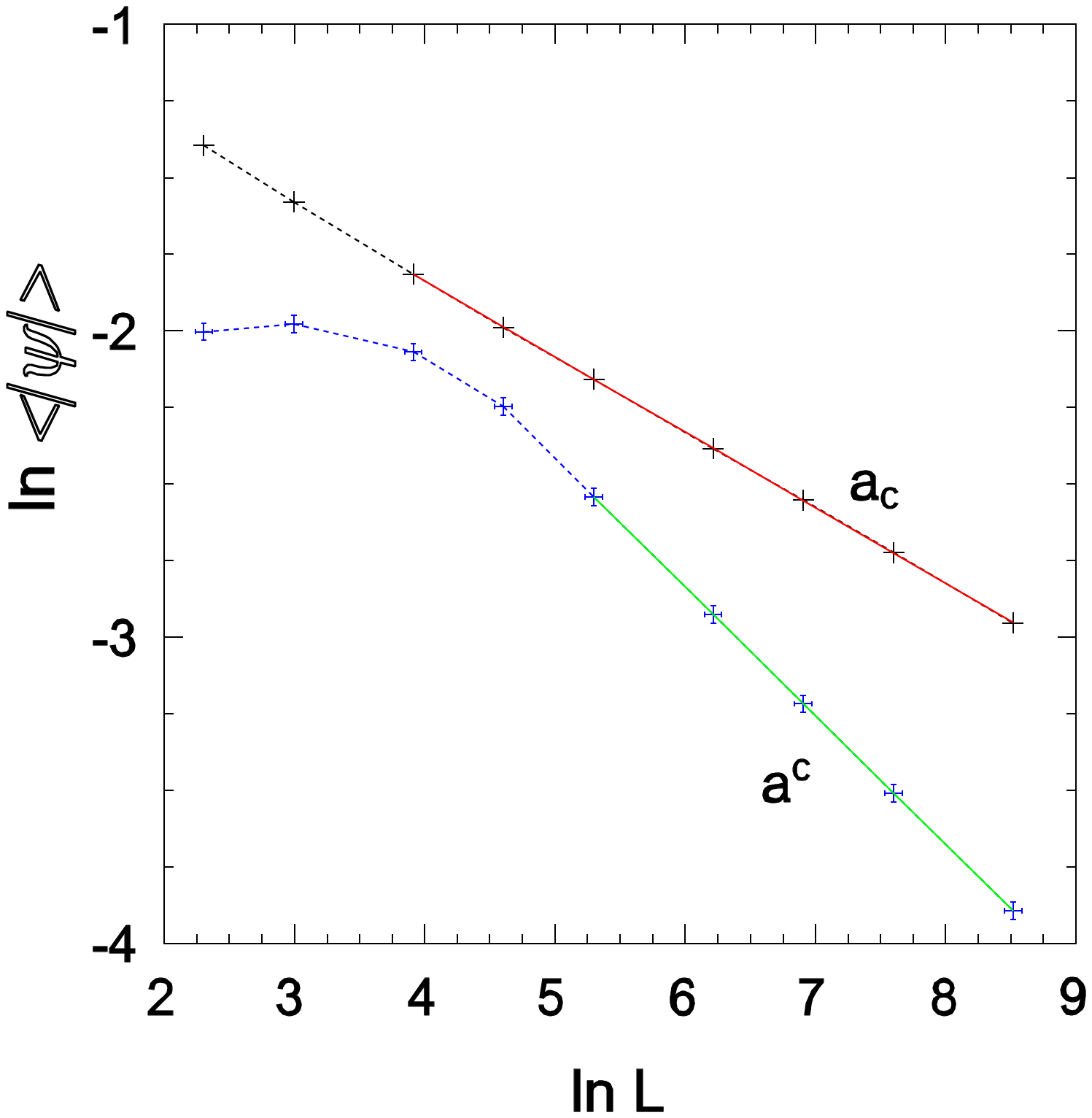}%
    \caption{\label{fig:psiL} $\langle |\psi| \rangle$ at the critical
      points versus system size~$N$ on the complete graph. The order
      parameter decays as a power law of the system size at both phase
      transitions.
       }%
  \end{center}%
\end{figure}%

Figure~\ref{fig:psi} shows the variation of~$\langle |\psi|
\rangle$~($\equiv \langle \langle |\psi| \rangle_t \rangle_s$)
with~$a$ in mean-field theory, and on the complete graph, confirming
that~$|\psi|$ functions as an order parameter to detect both phase
transitions. The~MF critical behavior is~$\langle |\psi| \rangle \sim
(a- a_c)^{1/2} $ for~$a \swarrow a_c$ and~$\langle |\psi| \rangle \sim
(a^c - a)^{1/2} $ for~$a \nearrow a^c$.  In the {\it pair
  approximation} for a lattice with coordination number~$z=6$ we
find~$a_c = 1.97579$ and~$a^c = 4.71727$; for~$z=8$ the corresponding
values are~$a_c = 1.80198$ and~$a^c = 4.10028$. In the pair
approximation, no transition is observed for~$z \leq 4$.

On the complete graph, the order parameter decays with system size
as~$\langle |\psi| \rangle \sim N^{-1/4}$ at~$a=a_c$ and
as~$N^{-0.4203(3)}$ at~$a=a^c$, as shown in Fig.~\ref{fig:psiL}. The
first result is typical of mean-field-like scaling with system size at
a continuous phase transition, and can be understood as follows. Our
numerical results show that with increasing~$N$, as expected, the
stationary probability distribution in the~$P_1-P_2$ plane becomes
(for~$a = a_c$) increasingly concentrated around the symmetric
point~$(1/3,1/3)$. The maximum of the distribution,~$p_{max}$, scales
as~$N^{-3/2}$ and the number of points at which the probability is
relatively large (i.e.,~$p(N_1,N_2) \geq p_{max}/2$) grows~$\sim
N^{3/2}$. In terms of the densities~$P_i = N_i/N$, the probability
distribution is concentrated on a region of radius~$\rho \sim
N^{-1/4}$. Note that~$|\psi|^2$ is a quadratic form of the
variables~$\xi_i \equiv P_i -1/3$ and that~$\psi=0$ for~$\xi_1 = \xi_2
= 0$. Thus for the purposes of scaling analysis we may write the
stationary average of~$|\psi|$ as
\begin{equation}
\langle|\psi|\rangle \sim \frac{\int_0^{N^{-1/4}} \xi^2 d
  \xi}{\int_0^{N^{-1/4}} \xi d \xi} \sim N^{-1/4}
\end{equation}
with~$\xi = \sqrt{\xi_1^2 + \xi_2^2}$.  At the first
transition~($a=a_c$) the same argument applies to the order
parameter~$r$, and we indeed find numerically~$r \sim N^{-0.26}$.

At the second transition, the probability distribution is peaked near
the three stable fixed points of the mean-field analysis, and so has a
very different form than that found at~$a_c$.  It is therefore not
surprising that~$\langle |\psi| \rangle$ decays with a different
exponent. The value of~0.4203(3) (which does not correspond to any simple
ratio of integers) nevertheless remains something of a puzzle.  We
suspect that understanding this result will require an
asymptotic~(large-$N$) analysis of the Fokker-Planck equation for the
probability density in the~$P_1-P_2$ plane, a task we defer to future
work.

\subsection{Hypercubic lattices }

Motivated by the appearance of a novel phase transition in mean-field
theory and on the complete graph, we search for such a transition on
finite-dimensional structures, i.e., the simple cubic lattice and its
four-dimensional~(hypercubic) analog, in systems of~$N = L^d$ sites.
On these lattices, we do observe a marked tendency for~$\omega$ to
fall rapidly with~$a$ for values well above the first transition
at~$a_c$~($a_c\simeq 2.345$,~$\simeq 1.900$ and~$\simeq 1.750$
for~$d=3$, 4 and 5, respectively~\cite{Wood06a}). We also observe a
broad maximum in~$\chi_r$ at some point in this range. To infer the
existence of a phase transition, rather than a rapid but smooth
variation of~$\omega$ and other properties with~$a$, we require
evidence of emergent singularities as the linear system size~$L$ is
increased.

\begin{figure}
\includegraphics[width=0.9\columnwidth,angle=0]{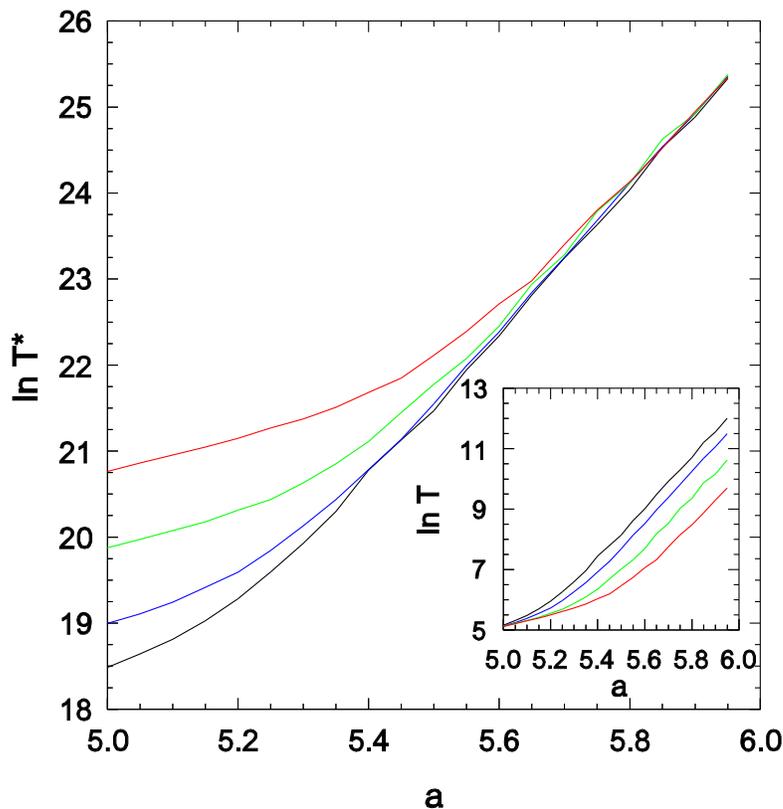}
\begin{center}
\caption{\label{T4D} (Color online)~Four dimensional hypercubic
  lattice: Inset mean period~$T$ versus a for~(upper to lower)~$L=28$,
  32, 40, and 50.  Main graph: scale period~$T^* \equiv L^d T$; system
  size increasing {\it upwards}.}
\end{center}
\end{figure}

The global dynamics may be characterized in terms of a process that
begins with all sites in state~$j$, and ends when the fraction of
sites remaining this state falls to~$\leq 1/3$. Let the ``escape
time"~$T$ denote the mean duration this process, averaged over
realizations of the stochastic dynamics. If there is an
infinite-period phase transition at some critical value~$a^c$,
then~$T$ should grow~$\sim L^z$ at this point, where~$z$ denotes the
dynamic critical exponent. Although simulations reveal an exponential
growth of~$T$ with~$a$ at fixed~$L$, they show that for large,
fixed~$a$, the escape time {\it decreases} with increasing system
size! The mean period (see Fig.~\ref{T4D}) exhibits two regimes for~$a
> a_c$: for smaller~$a$, relatively slow, exponential growth with~$T
\sim e^a$ essentially independent of system size, and, for larger~$a$,
rapid exponential growth with~$T \sim L^{-d} e^{\kappa a}$,
with~\mbox{$\kappa \gg 1$}. We identify the latter regime as that of
single-droplet nucleation~\cite{Rikvold94}. In the former regime, the
density of ``advanced" sites (i.e., in state~$j+1$ in a configuration
with the majority in state~$j$) is large enough that essentially no
barrier to the cyclic dynamics exists. The~(smooth) crossover between
these regimes occurs at a larger~$a$ value for larger system sizes.

The results for the mean transition rate~$T$ on the four-dimensional
hypercubic lattice,~$T \sim L^{-d}\exp[\kappa a]$, clearly suggest a
nucleation mechanism.  The exponential factor represents a barrier to
nucleation, that is, the mean time to formation of a critical cluster.
(By a ``critical" cluster we mean one that is equally likely to grow
as to diminish in size; larger clusters tend to grow while smaller
ones tend to shrink.) Consider a region in which all spins are in the
same state.  The rate at which a given spin flips to the next state
(e.g., 1~$\to$ 2) is~$e^{-a}$ leading to the exponential growth
in~$T$. The prefactor~$1/L^{d}$ corresponds to essentially independent
contributions due to each site in the system. (Given the extremely low
density of clusters in this regime, we can treat the nucleation and
initial growth of a cluster as occurring in isolation, without
interference from other clusters.)%
\begin{figure}%
\includegraphics[width=0.9\columnwidth,angle=0]{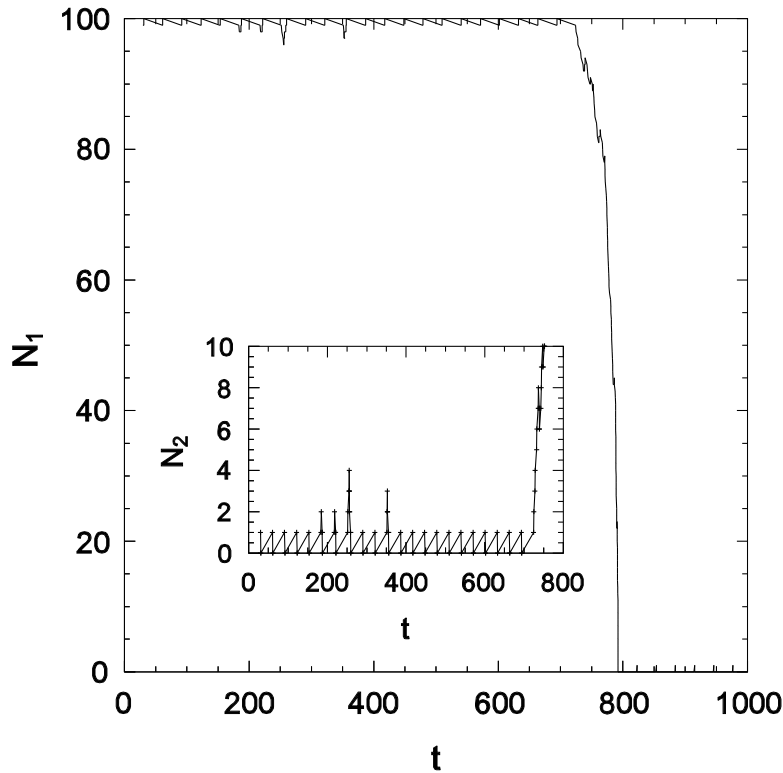}%
\begin{center}%
\caption{\label{N1A8} Square lattice,~$L=10$: evolution of~$N_1$~(main
  graph) and~$N_2$~(inset) in a typical realization with~$a=8$.}%
\end{center}%
\end{figure}%

The constant~$\kappa$ in the exponential reflects the probability of
formation of a critical cluster, which in turn depends on the critical
cluster size,~$\ell^*$.  Intuitively, the nucleation barrier exists
because an isolated advanced site has a higher rate to advance once
again, and then rapidly return to the majority state, than it does of
inducing a neighbor to advance.  Given the difficulty of visualizing
four-dimensional space, we begin by studying nucleation on the square
lattice.  Of course, the model does not exhibit an ordered phase in
two dimensions, but this issue can be bypassed by using small
systems~(a lattice of 10$\times$10 sites, in the present case).  Then
the system remains near one of the three attractors (almost all sites
in the same state) except for the rapid transitional periods following
a nucleation event.  (In larger systems global synchronization is
quickly lost and the system breaks into a number of domains.)  On such
a small lattice, one may easily identify transitional regimes between,
for example, states with majority-1 and majority-2.  We define the
transition (in the macroscopic sense) as the moment when 20\% of sites
are in state 1, with nearly all the rest (70\% or more) in state 2.
To estimate $T$ we determine the time required for a large number
(4000, in practice) of transitions to occur, in event-driven Monte
Carlo simulations.  A typical evolution is shown in Fig.~\ref{N1A8}.
We find that the mean transition time grows exponentially with $a$;
for $a \geq 7$ we find $\kappa = 1.68(1)$.  This suggests that the
critical cluster size is approximately 3 - 4.

\begin{figure}[!htb]%
\vspace{-0.5cm}%
\includegraphics[angle=0,width=\columnwidth]{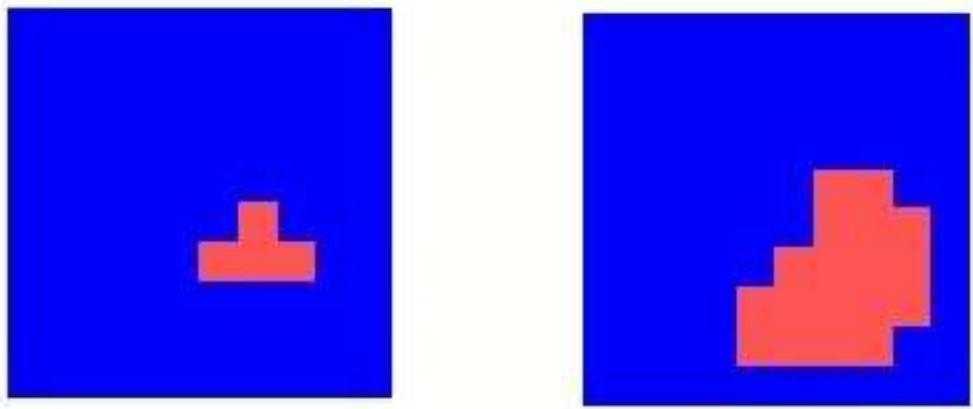}%
\caption{(Color online) Typical evolution of a cluster of 2s (red) in
  a background of 1s (blue).  Square lattice, $L=10$, $a=6$, times
  $t=39$ (left) and 58 (right).}%
\label{ev6}%
\end{figure}

Several other pieces of evidence support this estimate. First, in the
pre-transition regime, the number $N_2$ of sites in state 2 is seen to
fluctuate between zero and three or so.  In the example shown in
Fig.~\ref{N1A8}, the second time $N_2$ grows to four, a transition
occurs. Typical formats of a growing cluster are shown in
Fig.~\ref{ev6}; the cluster is seen to be compact, as expected for a
growth probability that is highest for the sites outside the cluster
having the largest number of neighbors within it.%

A semiquantitative analysis of the cluster size distribution in the
pretransition regime can be developed as follows.  Consider first the
density $\rho_2$ of isolated sites in state 2, in a background of 1s.
The rate for a given 1 to change to 2 is $e^{-a}$, while an isolated 2
(with all neighbors in state 1) flips to state 3 at unit rate.  There
are also transitions of the form 21 $\to$ 22.  The total rate for an
isolated 2 to become a dimer is $4 e^{-a/2}$.  Thus we have a rate
equation of the form
\begin{equation}
\dot{\rho}_2 = e^{-a} \rho_1 - \rho_2 - 4 e^{-a/2} \rho_1 \rho_2
\simeq e^{-a} - \rho_2
\end{equation}
leading to $\rho_2 \simeq e^{-a}$ in the quasi-stationary (QS) regime.
By the QS regime we mean the long-lived metastable state in which the
great majority of sites are in state 1. Similarly, let $\rho_{22}$
denote the probability that a given site is the leftmost (or
lowermost) site of an isolated dimer (the restriction is to avoid
double counting). Then we have
\begin{equation}
\dot{\rho}_{22} \simeq 4e^{-a/2} \rho_2 - 2 e^{-a/4} \rho_{22} - {\cal O}(e^{-a/2} \rho_{22})
\end{equation}
so that $\rho_{22} \simeq 2 e^{-5a/4}$ in the QS state.  Consider next
the density $\rho_{\sf L}$ of L-shaped trimers.  The gain term in the
equation for $\rho_{\sf L}$ is $4 e^{-a/2} \rho_{22}$ while the
dominant loss term (due to formation of a four-site, square cluster)
is simply $-\rho_{\sf L}$ (the transition rate is unity).  Thus
$\rho_{\sf L} \simeq 8 e^{-7a/4}$.  The exponential factor,
$e^{-7a/4}$, is quite close to the factor $e^{-1.68(1) a}$ found in
simulations, suggesting that this is essentially the critical cluster.
(Note that this is the smallest cluster with a larger probability of
growing than of shrinking.)  Applied to the simple cubic lattice, this
line of reasoning suggests that the critical cluster consists of seven
sites (in the form of a cube with one corner removed) and yields
$\kappa_3 = 11/3$.  [A crude estimate for the value in four dimensions
  can be obtained by supposing that $\kappa_4 \approx
  (\kappa_3/\kappa_2) \times \kappa_3$; this yields $\kappa_4 \approx
  7.7$, reasonably close to the simulation value, $\kappa_4 =
  8.43(7)$.]

The nucleation scenario outlined above essentially {\it rules out the
  second transition on any finite-dimensional lattice with short-range
  interactions}.  There is always a cluster size such that growth is
more likely than shrinkage, and once such a cluster forms, a
transition of the global state is likely.  From the perspective of
{\it equilibrium} phase transitions, this is not surprising: the
ordered state of any (finite) Ising or Potts model is always subject
to a reversal of magnetization at a finite rate.  (This rate, however,
decays with system size in equilibrium, as a finite cluster is always
more likely to shrink than to grow, in the absence of an external
field.)  In the present cyclic dynamics, by contrast, there is no
impediment to growth (no ``surface tension", as it were), once a
cluster has attained the critical size.

\section{\label{conclusions}Conclusions}

We have identified a second phase transition in the cyclic model
proposed by Wood et al. \cite{Wood06a,Wood06b,Wood07a,Wood07b} This
continuous transition is of a fundamentally different nature than the
first, as it corresponds to an infinite-period transition at which
macroscopic properties freeze and a discrete rotational ($C_3$)
symmetry is spontaneously broken, in the absence of an absorbing
state.  We find that an appropriate order parameter for the second
transition involves the local transition rates. While the phase
transition is observed on the complete graph, we argue that it does
not exist on finite-dimensional structures with short-range
interactions.  The question naturally arises whether such an
infinite-period transition exists in systems with long-range
interactions.  This indeed appears likely, for interactions that decay
sufficiently slowly with distance.  In the case of graphs with
interactions between many pairs of sites, independent of distance, we
expect to observe an infinite-period transition if (in the
infinite-size limit) a given site interacts with a finite fraction of
the others.  We plan to investigate these questions, as well as the
behavior on scale-free networks, in future work.

\section{Acknowledgments}
VRVA and MC acknowledge financial support from CNPq, FACEPE, CAPES,
FAPERJ and special programs PRONEX, PRONEM and INCEMAQ. RD
acknowledges financial support from CNPq.\\

\bibliography{copelli}

\end{document}